\documentstyle[aps,prl,multicol]{revtex}
\input{epsf}
\begin{document}
\draft
\preprint{\today}
\title{A new quantum phase between the Fermi glass and 
the Wigner crystal \\ in two dimensions}
\author{Giuliano Benenti, Xavier Waintal and Jean-Louis Pichard}
\address{CEA, Service de Physique de l'Etat Condens\'e,
           Centre d'Etudes de Saclay, 91191 Gif-sur-Yvette, France}
\maketitle
\begin{abstract} 
  For intermediate Coulomb energy to Fermi energy ratios $r_s$, 
spinless fermions in a random potential form a new quantum phase 
which is nor a Fermi glass, neither a Wigner crystal. Studying 
small clusters, we show that this phase gives rise to an ordered 
flow of enhanced persistent currents, for disorder strength and 
ratios $r_s$ where a metallic phase has been recently observed in 
two dimensions.
\end{abstract}
\pacs{PACS: 71.30.+h, 72.15.Rn}  

\begin{multicols}{2}
\narrowtext

%
%

 An important parameter for a system of charged particles is the Coulomb 
energy to Fermi energy ratio $r_s$. In a disordered two-dimensional 
system, the ground state is obvious in two limits.  For large $r_s$, 
the charges form a kind of pinned Wigner crystal, the Coulomb repulsion 
being dominant over the kinetic energy and the disorder. For small 
$r_s$, the interaction becomes negligible and the ground state is a Fermi 
glass with localized one electron states. There is no theory for 
intermediate $r_s$, while many transport measurements following the 
pioneering works of Kravchenko et al~\cite{kravchenko} and made with 
electron and hole gases give evidence of an intermediate metallic phase 
in two dimensions, observed~\cite{hamilton} for instance when 
$6 < r_s < 9$ for a hole gas in GaAs heterostructures. A simple model of  
spinless fermions with Coulomb repulsion in small disordered $2d$ 
clusters exhibits a new ground state characterized by an ordered flow 
of enhanced persistent currents for those values of $r_s$. In a given cluster, 
as we turn on the interaction, the Fermi ground state can be followed 
from $r_s=0$ up to a first level crossing. A second crossing 
occurs at a larger threshold after which the ground state can 
be followed to the limit $r_s \rightarrow \infty$. There is then an 
intermediate state between the two crossings. In small 
clusters, the location of the crossings depends on the considered  
potentials, but a study over the statistical ensemble of the currents 
supported by the ground state gives us two well defined values  
$r_s^{F}$ and $r_s^{W}$: Mapping the system on a torus threaded by an 
Aharonov-Bohm flux, we denote respectively $I_{l}$ and $I_{t}$ the total 
longitudinal (direction enclosing the flux) and transverse parts of the 
driven current. One finds for their typical amplitudes $|I_{t}| \approx 
\exp -(r_s/r_s^{F})$ and $I_{l} \approx \exp-(r_s/r_s^{W})$ with 
$r_s^{F} < r_s^{W}$. Below $r_s^F$, the flux gives rise to a glass 
of local currents and the sign of $I_{l}$ can be diamagnetic or 
paramagnetic, depending on the random potentials. Above $r_s^{F}$, 
the transverse current is suppressed while an ordered flow of longitudinal 
currents persists up to $r_s^{W}$, where charge crystallization occurs. 
The sign of $I_{l}$ can be paramagnetic or diamagnetic depending on the 
filling factor (as for the Wigner crystal), but does not depend on the 
random potentials (in contrast to the Fermi glass). One finds $r_s^{F}$ 
and $r_s^{W}$ in agreement with the values delimiting the new metallic 
phase when $ 0.3 < k_F l <3$, $k_F$ and $l$ denoting the Fermi wave vector 
and the elastic mean free path respectively. For $k_Fl \geq 1$, $I_l$ is 
strongly increased between $r_s^F$ and $r_s^W$. This suggests that 
the intermediate phase of our model is related to the new metal observed 
in two dimensions by transport measurement which we shortly review. 

%
%
 In exceptionally clean GaAs/AlGaAs heterostructures, an insulator-metal 
transition (IMT) of a hole gas results~\cite{yoon} from an increase  
of the hole density induced by a gate. This occurs at $r_s \approx 35$, 
in close agreement to $r_s^{W}\approx 37$, where charge crystallization 
takes place according to Monte Carlo calculations~\cite{tanatar}, 
and makes highly plausible that the observed IMT comes from the quantum 
melting of a pinned Wigner crystal. The values of $r_s$ where an IMT 
has been previously seen in various systems  (Si-Mosfet, Si-Ge, GaAS) 
are given in Ref.~\cite{yoon}, corresponding to different degrees of 
disorder (measured by the elastic scattering time $\tau$). Those $r_s$ 
drop quickly from $35$ to a constant value $r_s \approx 8-10$ when $\tau$ 
becomes smaller. This is again 
compatible with $r_s^{W} \approx 7.5$ given by Monte Carlo 
calculations~\cite{chui} for a solid-fluid transition in presence of 
disorder. If the observed IMT are due to interactions, it might be 
expected that this metallic phase will cease to exist as the carrier 
density is further increased. This is indeed the case~\cite{hamilton} 
for a hole gas in GaAs heterostructures at $r_s \approx 6$ where an 
insulating state appears, characteristic of a Fermi glass with 
electron-electron interactions.

%
%
   In this work, we take advantage of exact diagonalization 
techniques for large sparse matrices (Lanczos method) where tiny 
changes of energy can be precisely studied. This 
restricts us to small clusters and low filling factors. Fortunately, 
the dependence on particle number has proved to be remarkably weak 
in many cases. In the clean limit, calculations \cite{FQHE} with 
6-8 particles give the condensation of the electron gas into 
an incompressible quantum fluid when a magnetic field is applied. 
Pikus and Efros~\cite{tanatar} have obtained $r_s^{W}\approx 35$ from 
$6 \times 6$ clusters with $6$ particles, close to $r_s^{W}\approx 37$ 
obtained by Tanatar and Ceperley for the thermodynamic limit. In the 
disordered limit which we consider, there is another reason for 
expecting weak finite size effects. When the energy levels do not depend 
very much on the boundary conditions, the periodic repetition of the same 
cluster cannot drastically differ from the thermodynamic limit obtained 
from an ensemble of different clusters. This usual localization criterion 
applies for insulators as the Fermi glass or the pinned Wigner crystal. 
Small cluster approximations should then be sufficient for small and 
large $r_s$. This explains why the critical factors $r_s$ which we 
will discuss are close to the thermodynamic limit given by the 
experiments. Finite size effects can be important only if one has a 
metal for intermediate $r_s$.
%
%

 We consider a simple model of $N=4$ Coulomb interacting spinless fermions 
in a random potential defined on a square lattice with $L^2=36$ sites. 
The Hamiltonian reads: 
\begin{eqnarray} 
\label{hamiltonian} 
H=-t\sum_{<i,j>} c^{\dagger}_i c_j +  
\sum_i v_i n_i +U \sum_{i\neq j} \frac{n_i n_j}{2 r_{ij}}. 
\end{eqnarray} 
$c^{\dagger}_i$ ($c_i$) creates (destroys) an electron in 
the site $i$, $t$ is the strength of the hopping terms 
between nearest neighbors (kinetic energy) and $r_{ij}$ 
is the inter-particle distance for a $2d$ torus. The random potential 
$v_i$ of the site $i$ with occupation number $n_i=c^{\dagger}_i c_i$ 
is taken from a box distribution of width $W$. The interaction strength 
$U$ yields a Coulomb energy to Fermi energy ratio 
$r_s=U/(2t\sqrt{\pi n_e})$ for a filling factor $n_e=N/L^2$. 
The disorder to hopping energy ratio $W/t$ is chosen 
such that $k_F l$ takes values where the IMT has been observed 
\cite{kravchenko,hamilton,yoon}. A Fermi golden rule approximation for 
$\tau$ gives \cite{montambaux} $k_F l \approx 192 \pi n_e (t/W)^2$. One has 
$n_e=1/9$, $W/t=5, 10, 15$ corresponding to $k_Fl= 2.7, 0.67$ and 
$0.3$ respectively. 

 The boundary conditions are always taken periodic in the transverse 
$y$-direction, and such that the system becomes a torus enclosing an 
Aharonov-Bohm flux $\phi$ in the longitudinal $x$-direction. Imposing 
$\phi=\pi/2$ ($\phi=\pi$ corresponds to anti-periodic condition), one 
drives a persistent current of total longitudinal and transverse 
components given by 
\begin{equation}
I_{l}=- \frac{\partial E(\phi)}{\partial \phi}|_{\phi=\pi/2} 
=\frac{\sum_i I_i^l}{L}
\end{equation}
and $ I_{t}={\sum_i I_i^t}/{L}$ respectively. The 
local current $I_i^l$ flowing at the site $i$ in the longitudinal 
direction is defined by $I_i^l =2 {\rm Im} (\langle \Psi_0 | 
c^{\dagger}_{i_{x+1},i_y} c^{}_{i_x,i_y} | \Psi_0 \rangle)$ and by a 
corresponding expression for $I_i^t$. The response is paramagnetic 
if $I_{l}>0$ and diamagnetic if $I_{l} < 0$. We begin by showing 
behaviors characteristic of a single cluster when $r_s$ varies.

\begin{figure}
\centerline{
\epsfxsize=9cm 
\epsfysize=11cm 
\epsffile{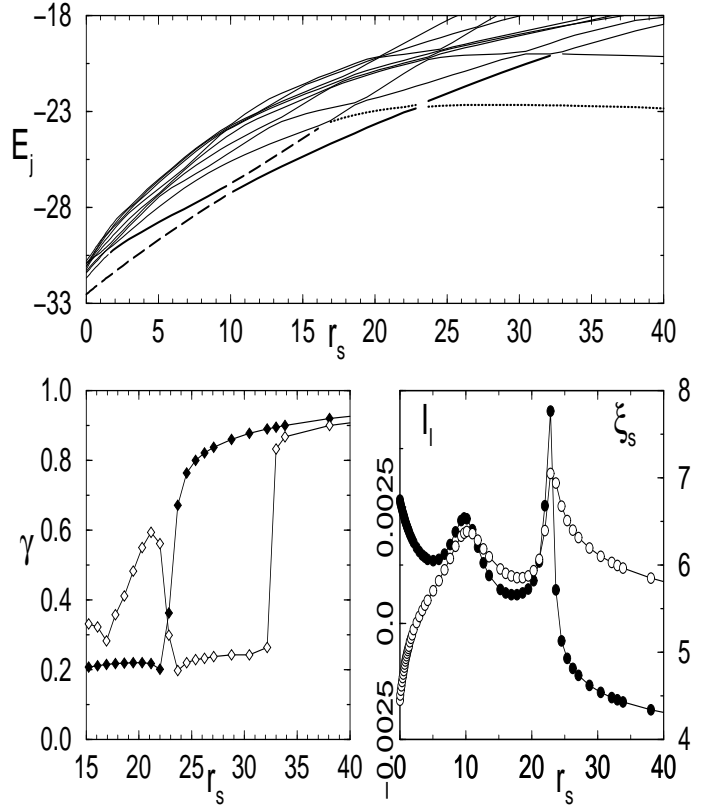}
}
\caption{ 
Behavior of a single cluster for $k_Fl \leq 1$ as a function of $r_s$. 
Top: Low energy spectrum (a $1.9 r_s$ term has 
been subtracted); Bottom left: jumps of $\gamma$ at the second crossing 
where the ground state (filled diamonds) and the first excited state (empty 
diamonds) are interchanged. Bottom right: $I_{l}$ (left scale, empty 
circles) and number of occupied sites $\xi_s$ (right scale, filled circles).
}
\label{fig1} 
\end{figure}

%
 Fig.~\ref{fig1} corresponds to $k_fl \leq 1$ ($W/t=15$). Looking at the 
low energy part of the spectrum, one can see 
that, as we gradually turn on the interaction, classification of the levels 
remains invariant up to first avoided crossings, where a Landau theory 
of the Fermi glass is certainly no longer possible. Looking at the 
electronic density $\rho_i = \langle \Psi_0| n_i| \Psi_0 \rangle$ of 
the ground state $|\Psi_0\rangle$, we have checked that it is mainly 
maximum in the minima of the site potentials for the Fermi glass. After the 
second avoided crossing, $\rho_i$ is negligible except for four sites 
forming a lattice of charges as close as possible to the Wigner crystal 
triangular network in the imposed square lattice. The degeneracy of the 
crystal is removed by the disorder, the array being pinned in $4$ sites 
of favorable energies.  
For the same cluster, we have calculated 
$C(r)=N^{-1} \sum_i \rho_i \rho_{i-r}$ and the parameter 
$\gamma=\max_{\,r} C(r) - \min_{\,r} C(r)$ used by Pikus and 
Efros~\cite{tanatar} for characterizing the melting of the crystal. 
$\gamma=1$ for a crystal and $0$ for a liquid. Calculated 
for the ground state and the first excited state, $\gamma(r_s)$ 
allows us to identify 
the second crossing with the melting of the crystal. Moreover, one can 
see that the crystal becomes unstable in the intermediate phase, while 
the ground state is related to the first excitation of the crystal 
(Fig.~\ref{fig1} bottom left). Around the crossings, the longitudinal 
current $I_l$ and the participation ratio $\xi_s =N^2( \sum_i \rho_i^2)^{-1}$ 
of the ground state (i.e. of the number of sites that it occupies) are 
enhanced (Fig.~\ref{fig1} bottom right). The general picture 
is somewhat reminiscent of strongly disordered chains~\cite{sjwp} where 
level crossings associated to charge reorganizations 
of the ground state are accompanied by enhancements of the persistent 
currents. Fig.\ref{fig1} is representative of the ensemble, with the 
restriction that the location of the crossings fluctuates from one sample 
to another as well as the sign (paramagnetic or diamagnetic) of $I_l$ 
below the first crossing, in contrast to $1d$.
\begin{figure}
\centerline{
\epsfxsize=8cm 
\epsfysize=7cm 
\epsffile{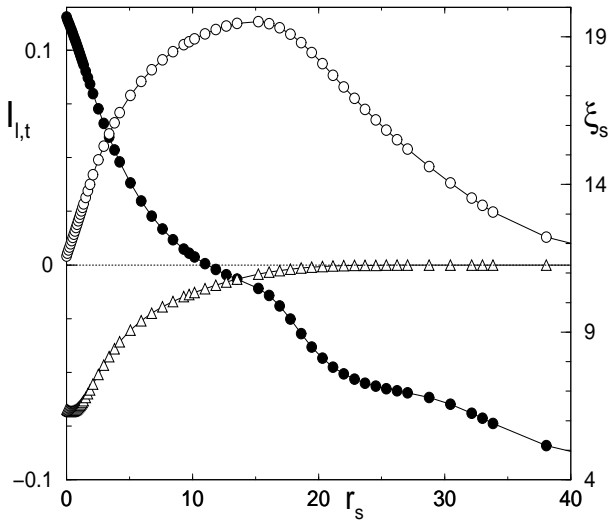}
}
\caption
{ 
Behavior of a single cluster for $k_Fl \geq 1$ as a function of $r_s$. 
Left coordinates: currents $I_l$ (circle) and $I_t$ (triangle). 
Right coordinates: $\xi_s$ (filled circle). 
}
\label{fig2} 
\end{figure} 
 Fig.~\ref{fig2} corresponds to $k_fl \geq 1$ ($W/t=5$). 
The previous level crossings are now almost suppressed by a stronger 
level repulsion and charge crystallization occurs more continuously.  
There is instead a broad enhancement of $I_l$ which, in contrast to 
Fig. \ref{fig1}, is not accompanied by a corresponding increase of 
$\xi_s$, which smoothly decreases from $20$ of the $36$ possible sites 
down to $4$ when charge crystallization becomes perfect. A transition of the 
persistent current, from a disordered array of loops towards an ordered flow 
as $r_s$ increases has been noticed~\cite{avishai} by Berkovits and Avishai. 
To illustrate this phenomenon, the total transverse current $I_t$ is 
shown in Fig. \ref{fig2}. One can see that $I_t$ is suppressed at 
$r_s \approx 5$ while $I_l$ continues to increase up to $r_s \approx 15$. 
We have checked that a disordered array of loops persists up to $r_s 
\approx 5$, followed by an ordered flow of enhanced longitudinal currents 
persisting up to $r_s \approx 15$. The disordered array of loops gives 
rise to a diamagnetic or paramagnetic current $I_l$, depending on the 
microscopic disorder. The ordered flow gives rise to a paramagnetic $I_l$. 
However, Coulomb repulsions do not always yield a paramagnetic response. 
For instance, $4\times6$ clusters with $N=6$ become always diamagnetic at 
large $r_s$. One can only conclude that the sign of the response in $2d$ 
does not depend on the random potential when $r_s$ is sufficient for 
suppressing $I_t$. 
%
%
 In $1d$, Legett's theorem~\cite{legett} states that the sign of 
$I_l$ depends on the parity of $N$ only, for all disorder and interaction 
strength. The proof is based on the nature of ``non symmetry dictated nodal 
surfaces '', which is trivial in $1d$, but which has a quite complicated 
topology in higher $d$. It is likely that such a theorem could be extended 
in $2d$ when the transverse flow is suppressed. 
%
%

\begin{figure}
\centerline{
\epsfxsize=9cm 
\epsfysize=11cm 
\epsffile{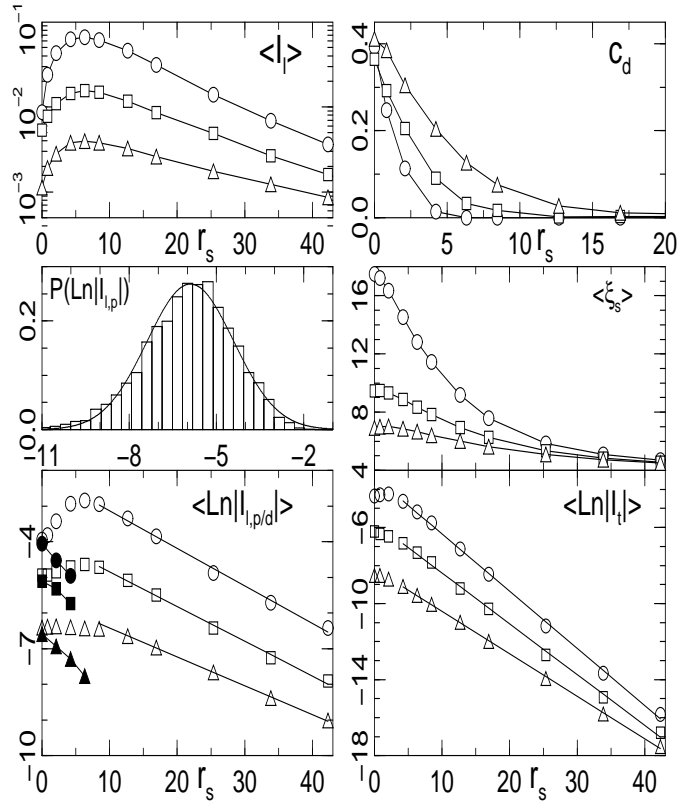} 
}
\caption{Statistical study of an ensemble of clusters for 
$W/t= 5$, (circle) $10$ (square) and $15$ (triangle) as a 
function of $r_s$. Top left: mean value $<I_l>$;  
Top right: fraction $C_d$ of diamagnetic samples; 
Middle left: distribution of the logarithms of the paramagnetic 
current $I_{l,p}$ at $r_s=1.7$ and $W/t=15$. Middle right: 
mean number of occupied sites by the ground state. Bottom left: 
longitudinal paramagnetic (empty symbols) and diamagnetic (filled symbols) 
currents. Bottom right: transverse currents. The straight lines 
are exponential fits giving $r_s^F$ and $r_s^W$ shown in Fig. \ref{fig4}.
}
\label{fig3} 
\end{figure}
\begin{figure}
\centerline{
\epsfxsize=8cm 
\epsfysize=6cm
\epsffile{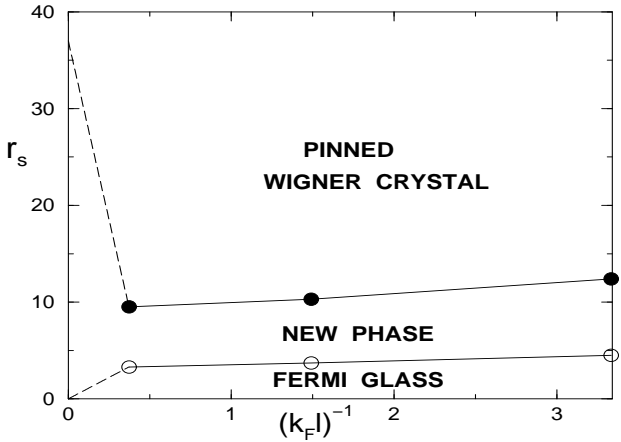} 
}
\caption{Proposed phase diagram for $2d$ spinless fermions in a 
random potential. $r_s^W$ (filled circle) and $r_s^F$ (empty circle) 
obtained from Fig. \ref{fig3} bottom.
}
\label{fig4} 
\end{figure}

 
 We now present a statistical study of an ensemble of $10^3$ clusters 
for $W/t=5,10,15$. At the left top of Fig. \ref{fig3}, 
one can see an increase of the mean $I_l$ by about one order of magnitude 
when $r_s \approx 7$ for $W/t=5$. We note that the persistent 
currents \cite{current} measured in an ensemble of mesoscopic rings are 
typically higher by a similar amount than the theoretical prediction 
neglecting the interactions. At the right top of Fig. \ref{fig3}, the 
fraction of diamagnetic clusters is given as function of $r_s$, 
showing that the enhancement of the mean is partially related to the 
suppression of the diamagnetic currents. 
This suppression is faster for weak disorders. The mean number 
$\xi_s$ of sites occupied by the ground state is given at the middle right 
of Fig. \ref{fig3}, showing a negligible increase when $W/t=15$ at low 
$r_s$ and a regular decay otherwise. The paramagnetic $I_{l,p}$ and 
diamagnetic $I_{l,d}$ longitudinal currents, and $|I_t|$ have log-normal 
distributions for all values of $r_s$ when $W/t \geq 5$. The stronger is 
the disorder, the better is the log-normal shape of the distribution 
(see middle left of Fig. \ref{fig3}). The average of the logarithms give 
the typical values shown in the bottom part of Fig. \ref{fig3}. On the left, 
the longitudinal currents $I_l$ are given, the diamagnetic responses $I_{l,d}$ 
(filled symbols) being separated from the paramagnetic responses $I_{l,p}$ 
(empty symbols), while the transverse currents $I_t$ are given at the right 
side. The log-averages exponentially decay as $I_{l,d} \propto |I_t| 
\propto \exp-(r_s/r_s^{F})$ and $I_{l,p} \propto \exp-(r_s/r_s^{W})$ 
when $r_s$ is large enough. The variances of $\log |I_{t}|$ and 
$\log I_{l}$ increase as $r_s/r_s^{F}$ and $r_s/r_s^{W}$ above 
$r_s^F$ and $r_s^{W}$ respectively. The values of $r_s^F$ and $r_s^W$ 
extracted from the exponential fits (straight lines of Fig. \ref{fig3}) 
are given in Fig. \ref{fig4}, where a sketch of the phase diagram 
is proposed.  
%
%
 Fig. \ref{fig3} and Fig. \ref{fig4} show that a simple model of spinless 
fermions with Coulomb repulsion in a random potential can account for the 
critical carrier densities and disorder strengths where the IMT occurs. 
The comparison between the curve $r_s (\tau)$ given in Ref. \cite{yoon} 
(summarizing the factors $r_s$ where the IMT has 
been observed) and the curve $r_s^W (k_Fl)$ of Fig. \ref{fig4} 
(characterizing the suppression of $I_l$) is very striking. The 
value $r_s=6$ where the reentry has been observed in Ref. \cite{hamilton} 
is also compatible with the curve $r_s^F (k_Fl)$ characterizing the 
suppression of $I_t$. We have not indicated in the proposed phase 
diagram the difference between $k_Fl \geq 1$ (where $I_l$ has a strong 
enhancement which may be the signature of a new metal at the 
thermodynamic limit) and $k_Fl \leq 1$ (where 
$I_l$ persists up to $r_s^W$ without noticeable enhancement). A study of the 
size dependence at a fixed filling factor will be necessary for studying 
a possible IMT driven by an increase of $k_Fl$ at intermediary $r_s$. 
This is unfortunately out of reach of exact diagonalization techniques. 
Another striking difference is that $\xi_s$ and $I_l$ convey similar 
information when $k_Fl <1$ while the increase of $I_l$ is accompanied by a 
decrease of $\xi_s$ when $k_Fl > 1$. This suggests that transport for 
intermediary $r_s$ results more from a collective motion of charges 
than from a delocalization of individual charges.
 The spin degrees of freedom are not included in our model, the orbital part 
of the wave function being totally anti-symmetrized. This restriction is 
quite important for short range screened interactions, but is certainly 
less severe for long range interactions and low densities. However, there 
are many experimental evidences \cite{cond-mat} that spin effects play a 
role.  A more complex phase diagram is possible when spins are included. 
But even for $2d$ spinless fermions, these small cluster studies allow us 
to conclude that there is a new quantum phase, clearly separated from 
the Fermi glass and from the Wigner crystal, identified by a plastic flow 
of currents without charge crystallization, which is likely to give a new 
metal in the thermodynamic limit when $k_Fl \geq 1$. 

 This work is partially supported by a TMR network of the EU.

\end{multicols}
\end{document}